\documentclass[aps,prl,twocolumn,reprint,showpacs,groupedaddress]{revtex4-1}
\usepackage{amsmath,amssymb,graphicx,color}
\usepackage{times}
\usepackage{latexsym}
\usepackage{stmaryrd}
\usepackage[table]{xcolor}
\usepackage[colorlinks,citecolor=blue]{hyperref}
\usepackage{amsfonts}
\usepackage{bm}
\usepackage{bbm}
\usepackage{color}
\usepackage[english]{babel}
\usepackage{stmaryrd}
\usepackage{psfrag,graphicx}
\usepackage{subfigure}
\usepackage{natbib}
\usepackage{appendix}

\newcommand\ket[1]{\left|\textstyle{#1}\right\rangle}
\newcommand\bra[1]{\left\langle\textstyle{#1}\right|}
\newcommand\braket[1]{\left\langle\textstyle{#1}\right\rangle}
\newcommand\half{\frac{1}{2}}
\newcommand\down{\downarrow}
\newcommand\up{\uparrow}

\begin{document}
\title{Quantum Phase Transition in the Finite Jaynes-Cummings Lattice Systems}
\author{Myung-Joong Hwang and Martin B. Plenio}
\affiliation{Institut f\"{u}r Theoretische Physik and IQST, Albert-Einstein-Allee 11, Universit\"{a}t Ulm, D-89069 Ulm, Germany}

\begin{abstract}
Phase transitions are commonly held to occur only in the thermodynamical limit of
large number of system components. Here we exemplify at the hand of the exactly solvable
Jaynes-Cummings (JC) model and its generalization to finite JC-lattices that finite
component systems of coupled spins and bosons may exhibit quantum phase transitions (QPT).
For the JC-model we find a continuous symmetry-breaking QPT, a photonic condensate with
a macroscopic occupation as the ground state and a Goldstone mode as a low-energy excitation.
For the two site JC-lattice we show analytically that it undergoes a Mott-insulator to superfluid
QPT. We identify as the underlying principle of the emergence of finite size QPT the
combination of increasing atomic energy and increasing interaction strength between the
atom and the bosonic mode which allows for the exploration of an increasingly large
portion of the infinite dimensional Hilbert space of the bosonic mode. This suggests that
finite system phase transitions will be present in a broad range of physical systems.
\end{abstract}
\maketitle

{\it Introduction.---} Quantum phase transition (QPT) and spontaneous symmetry
breaking are fundamental concepts in physics that lie at the heart of our
understanding of various aspects of nature, e.g., phases of matter such as
magnetism and superconductivity~\cite{Sachdev:2011uj,Sondhi:1997kc} or the generation
of mass~\cite{Anderson:1963gh,Higgs:1964gu} in high energy physics. A second-order
QPT is characterised by a closing spectral gap and degenerate ground states
with a spontaneously broken symmetry. A QPT is typically held to occur only
in the thermodynamical limit, i.e. a system with a diverging number of constituent
particles or lattice sites~\cite{Sachdev:2011uj}. A finite system size
generally opens the spectral gap, lifts the ground state degeneracy and restores
the symmetry of the ground state~\cite{Fisher:1972tv, Botet:1983bf}.

A notable exception is a recent finding in Ref.~\cite{Hwang:2015eq} concerning the
Rabi model~\cite{Braak:2011hc,Hwang:2010jn,Ashhab:2010eh,Ashhab:2013ke,Larson:2007fi},
which describes a single-mode cavity field coupled to a two-level atom. While the
Dicke model, a $N$-atom generalization of the Rabi model, has long been known for having
a QPT for $N\rightarrow\infty$~\cite{Hepp:1973jt,*Wang:1973ky,Emary:2003da}, Ref.~\cite{Hwang:2015eq}
demonstrates that the Rabi model itself undergoes a QPT with the same universal properties
when the ratio $\eta$ of the transition frequency to the cavity frequency diverges \cite{Ashhab:2013ke,Bakemeier:2012ja,Klimov:2009uv}. The study~\cite{Hwang:2015eq} further 
corroborates the duality between the system size $N$ and the frequency ratio $\eta$ by 
showing that the finite-size scaling exponents for $\eta$ are identical to those for $N$~\cite{Vidal:2006ex,Botet:1982ju}. It is then urgent and important task to see if
reaching a limit of the QPT for a system of finite components is a principle that
is generally applicable to photonic (phononic) systems with different underlying
symmetries, phases, and dimensions. If positively answered, it could open up an
important possibility of experimentally investigating the critical phenomena, both
in and out of equilibrium, in a small and fully-controlled quantum system.

In this letter, we consider the Jaynes-Cummings (JC) model~\cite{Jaynes:1963fa}, the Rabi model without
the so-called counter-rotating terms, which due to its $U(1)$ symmetry is
exactly solvable. We first point out that the well-known analytical solution of the
JC model exhibits a ground state instability in the $\eta\rightarrow\infty$
limit, in the sense that the ground state can lower its energy indefinitely by increasing
its photon occupation. In this unstable regime, we derive the analytical solution for the ground
state and the excitation spectrum by developing a low-energy effective theory. It shows
that the JC model undergoes a second order superradiant QPT  in the $\eta\rightarrow\infty$
limit. In the broken-symmetry phase, we find that the ground state forms a photon condensate
with a macroscopic photon occupation number and that the excitation spectrum is gapless
because the Goldstone mode~\cite{Goldstone:1962ff} emerges due to the broken continuous symmetry.

We develop this further by showing that the JC lattice model with only two lattice
sites, the JC dimer, undergoes a Mott-insulating-superfluid QPT in the same $\eta\rightarrow\infty$
limit. While the JC lattice model has been known to undergo a Mott-insulating-superfluid
QPT in the limit of infinite lattice sites \cite{Hartmann:2006kv,Greentree:2006jg,Angelakis:2007ho,Hartmann:2008ei,Koch:2009hh,Rossini:2007fx},
here, the QPT in the $\eta\rightarrow\infty$ limit is supported by the infinite
dimensional Hilbert space associated with the harmonic oscillator degree of freedom. Our exact analytical
solution of the ground state and the excitation spectrum shows that (i) the anti-symmetric normal
mode of the coupled-cavities undergoes a transition from a zero excitation insulating phase to a
superfluid phase with a broken global $U(1)$ symmetry, while the symmetric mode gets merely squeezed
in the superfluid phase and (ii) that the spectral gap of the anti-symmetric mode closes at the
critical point, beyond which the excitation is gapless, while the symmetric mode remains gapped
for any coupling strength. We emphasize that our analysis is analytic and fully quantum
mechanical, going beyond the mean-field approach that is often used in the studies of the JC
lattice model for the lack of the exact methods~\cite{Greentree:2006jg,Koch:2009hh}.

{\it Quantum phase transition in the JC model.---}
The Jaynes-Cummings Hamiltonian reads
\begin{equation}
\label{eq:1:01}
H_\textrm{JC}=\omega_0 a^\dagger a +\frac{\Omega}{2}\sigma_z-\lambda (a \sigma_++a^\dagger \sigma_-).
\end{equation}
Here, $\sigma_{\pm}=(\sigma_x\pm i \sigma_y)/2$ with $\sigma_{x,y,z}$ being the Pauli
matrices, and $a$ and $a^\dagger$ are the lowering and raising operator of a single mode
cavity field, respectively. The cavity frequency is $\omega_0$, the transition frequency of
the two-level atom $\Omega$ , and the coupling strength $\lambda$. The conserved total
number of excitation, $N_\textrm{tot}=a^\dagger a +\sigma_+\sigma_-$, leads to a $U(1)$-continuous
symmetry, that is, the model is invariant under a gauge transformation $U_\theta=e^{i\theta N_\textrm{tot}}$.
Let us denote $\ket{n}$ as a $n$-photon Fock state and $\ket{\up(\down)}$ as an eigenstate of
$\sigma_z$ with an eigenvalue $1(-1)$. We also introduce a dimensionless coupling strength
$g=\lambda/\sqrt{\omega_0\Omega}$ and a frequency ratio $\eta=\Omega/\omega_0$. Typically,
the Jaynes-Cummings model is obtained as an approximation to the Rabi model by neglecting the
so-called counter rotating terms, $-\lambda(a \sigma_-+a^\dagger \sigma_+)$~\cite{Hwang:2015eq}.
In systems where the atom-field interaction can be engineered, such as in a circuit
QED~\cite{Baksic:2014ih}, trapped-ion systems~\cite{Cirac:1993db} such counter-rotating terms
can be strongly suppressed and, remarkably, for an atomic $\Delta m=\pm 1$ transition in interaction
with circularly polarized light mode the rotating wave approximation is exact such that Eq.~(\ref{eq:1:01}) becomes a
correct description for any $g$~\cite{Crisp:1991eu}.

\begin{figure}[t]
\centering
\includegraphics[width=0.44\linewidth,angle=0]{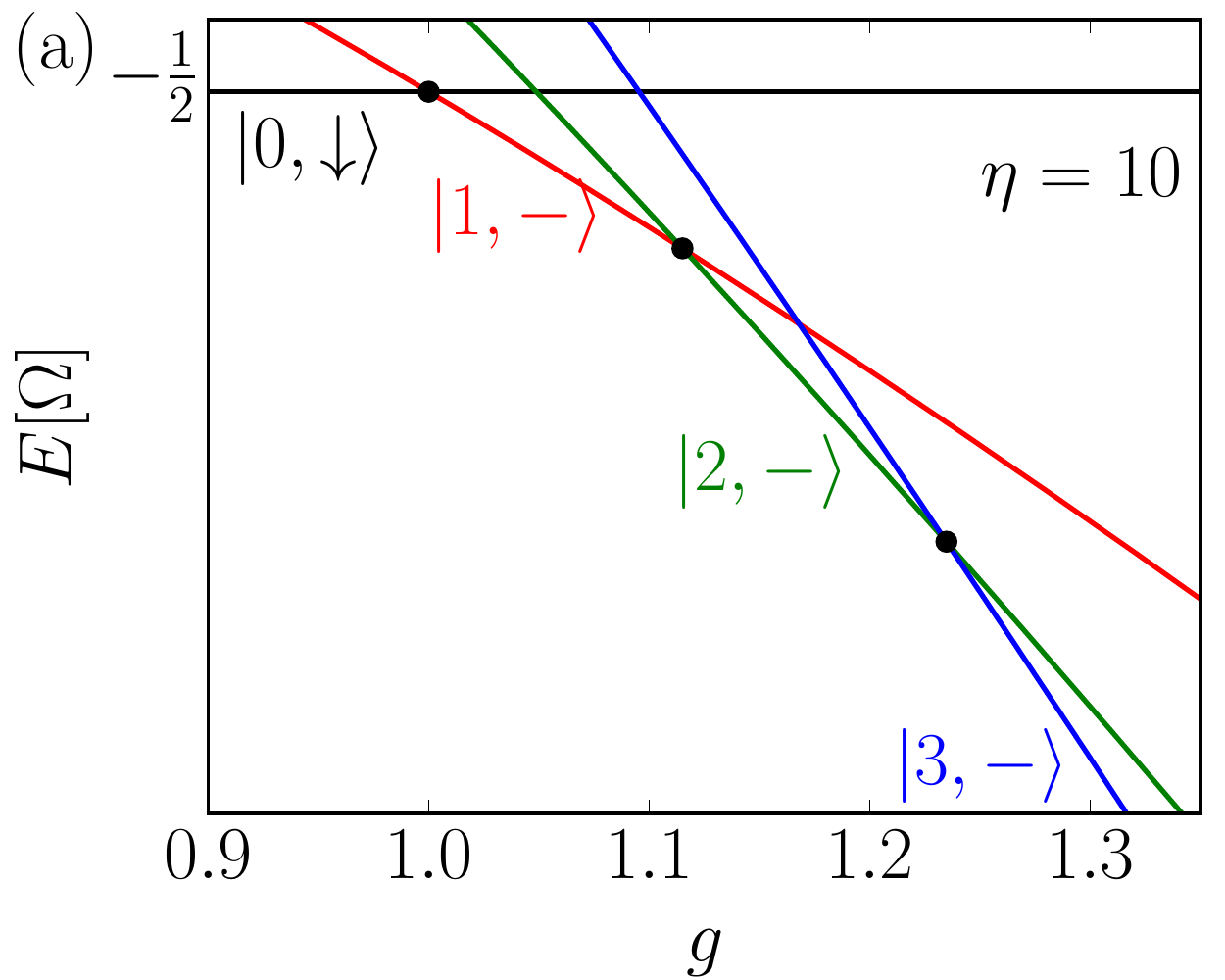}
\includegraphics[width=0.44\linewidth,angle=0]{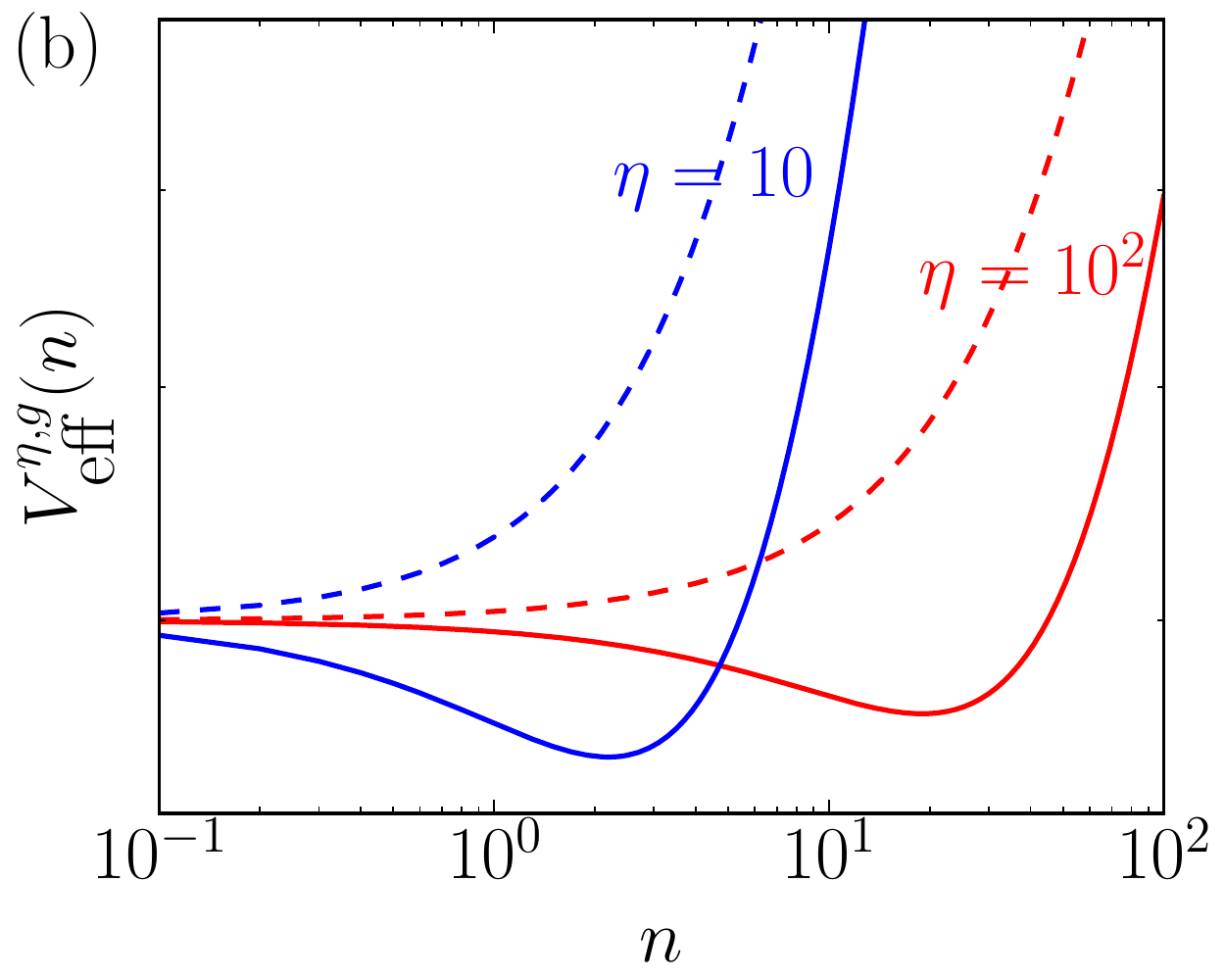}
\includegraphics[width=0.95\linewidth,angle=0]{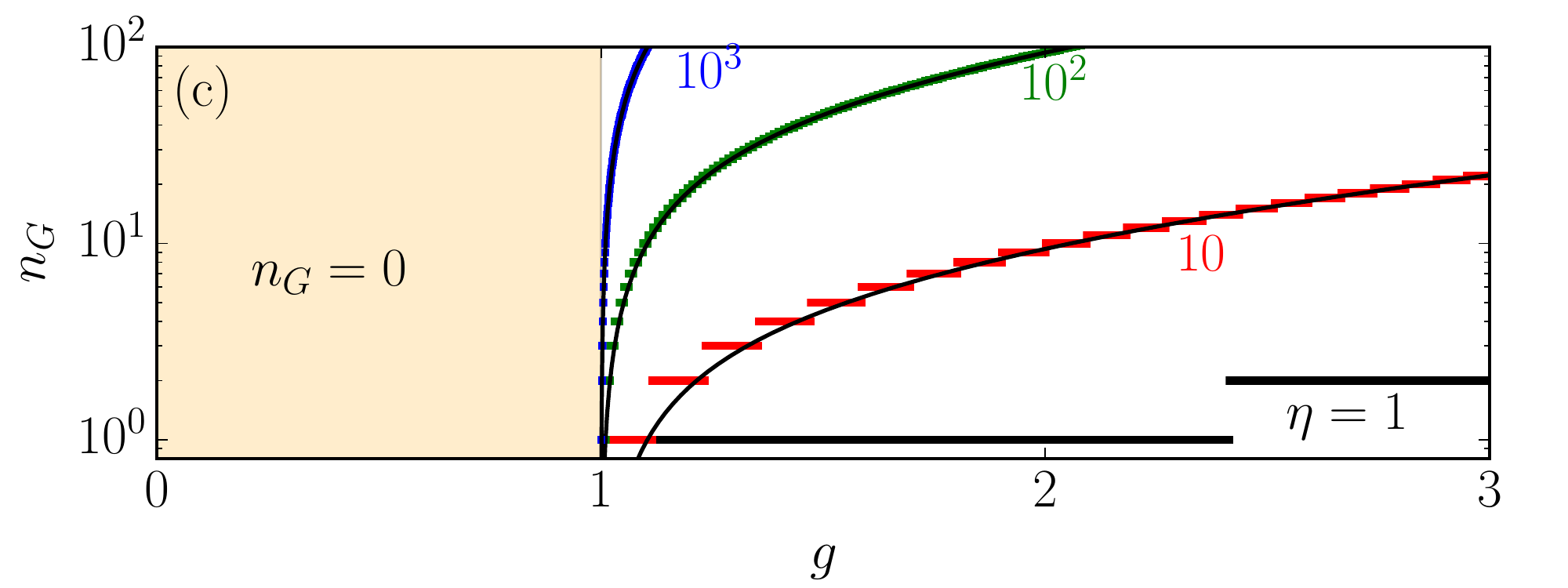}\\
\caption{Analytic solution of the JC model. (a) Level crossings for the ground state for a frequency ratio $\eta=10$. (b) An effective potential $V_\textrm{eff}^{\eta,g}$ for $g=0.8$ (dashed) and $g=1.2$ (solid) for different values of $\eta=10$ and $100$. (c) The total number of excitation of the ground state. As $\eta$ increases, the change near $g=1$ becomes progressively sharper, which is well described by $n_\textrm{sp}(g)=\eta(g^2-g^{-2})/4$ (solid).}
\label{fig:1}.
\end{figure}

The vacuum state $\ket{0,\down}$ is an energy eigenstate of the JC model with an eigenvalue
$E_{0,\down}=-\Omega/2$. There are two basis states with $n$ total number of excitation,
$\ket{n,\down}$ and $\ket{n-1,\up}$, which span the so-called JC-doublet, denoted as
$\ket{n,\pm}$, whose energy eigenvalues in unit of $\Omega$ read
\begin{equation}
    \label{eq:1:02}
    E_{n,\pm}(\eta,g)/\Omega= (n-\frac{1}{2})\eta^{-1}\pm\frac{1}{2}\sqrt{(1 -\eta^{-1})^2+4 g^2 n\eta^{-1} }.
\end{equation}
Regardless of $\eta$, for $g<1$ the ground state of Eq. (\ref{eq:1:01}) is always
$\ket{0,\down}$, until at $g=1$ there occurs a level crossing between $\ket{0,\down}$
and $\ket{1,-}$. This is followed by a series of level crossings between the lower-energy
states of adjacent JC doublets, $\ket{n,-}$ and $\ket{n+1,-}$ [Fig.~\ref{fig:1} (a)].
Therefore, increasing the atom-cavity coupling strength increases $\braket{N_\textrm{tot}}$
in the ground state, $n_G$, in discrete steps [Fig.~\ref{fig:1} (c)]; in this sense,
the JC-type atom-cavity coupling itself assumes the role of a chemical potential. Moreover,
as $\eta$ increases, the increase of $n_G$ becomes progressively sharper near at $g=1$
[Fig.~\ref{fig:1} (c)].

In the $\eta\rightarrow\infty$ limit, the nonlinearity in the spectrum of the JC model
disappears, leading to a harmonic spectrum in the low energy sector, i.e., $\lim_{\eta\rightarrow\infty}(E_{n,-}(\eta,g)-E_{0,\down})= \omega_0(1-g^2)n$, which is
a valid expression for any finite $n$. The excitation energy for $g<1$, a normal phase,
is therefore $\epsilon_\textrm{np}=\omega_0(1-g^2)$, which becomes zero at $g=1$, leading
to a degeneracy between $\ket{n,-}$ of any finite $n$ and $\ket{0,\down}$. For $g>1$,
Eq.~(\ref{eq:1:02}) shows a ground state instability in a sense that the ground state
energy can be indefinitely lowered by increasing $n$, until $n$ becomes infinite, where
a new energy minimum can be found. Regarding the Eq.~(\ref{eq:1:02}) as an effective
potential for photon numbers, $V^{\eta,g}_\textrm{eff}(n)$, for a fixed value of $\eta$
and $g$, we find the potential minimum at $n=0$ for $g<1$ and for $n>0$ for $g>1$ and
any $\eta$ [Fig.~\ref{fig:1} (b)]. For $\eta\gg1$, the potential minimum is located at
$n_\textrm{sp}(g)\equiv n_G(g>1)=\eta(g^2-g^{-2})/4+\mathcal{O}(\eta^0)$, which explains
very well the quadratic behavior of $n_G$ shown in Fig.~\ref{fig:1} (c). Furthermore,
in the $\eta\rightarrow\infty$ limit and $g>1$ it is immediate that $n_G$ diverges; that
is, a ground state superradiance occurs.

The instability of the JC model for $g>1$ in the $\eta\rightarrow\infty$ limit
predicted from Eq.~(\ref{eq:1:02}) and the finite value of $n_{sp}$ suggests the
derivation of a low-energy effective Hamiltonian for the superradiant phase $g>1$
that is valid around the potential minimum. To this end, we displace the cavity
field $a$ in Eq.~(\ref{eq:1:01}) by a complex number $\alpha=\alpha_ge^{i\theta}$
with $\alpha_g=\sqrt{n_\textrm{sp}}=\sqrt{\eta (g^2-g^{-2})/4}$, i.e., $\bar H_\textrm{JC}(\alpha_g,\theta)=\mathcal{D}^\dagger[\alpha] H_\textrm{JC}\mathcal{D}[\alpha]$.
By factoring out the phase, we have
\begin{align}
    \label{eq:1:03}
    \bar H_\textrm{JC}(\alpha_g)&=e^{-i\theta N_\textrm{tot}}\bar H_\textrm{JC}(\alpha_g,\theta)e^{i\theta N_\textrm{tot}}\nonumber\\
    &=\omega_0(a^\dagger a+\alpha_g^2)-\frac{\omega_0\sqrt{\eta}}{2g}(x\tau_x-g^2p\tau_y)+\frac{g^2\Omega}{2}\tau_z\nonumber\\
    &+\omega_0\alpha_gx(\tau_0+\tau_z)
\end{align}
Here we introduce the new spin operators $\tau_z=\ket{\bar \up}\bra{\bar \up}-\ket{\bar \down}
\bra{\bar \down} = g^{-2} \sigma_z - \sqrt{1-g^{-4}}\sigma_x$ as well as $x=a+a^\dagger$ and
$p=i(a^\dagger -a)$. Note that $\bar H_\textrm{JC}(\alpha_g)$ no longer
possesses the $U(1)$ symmetry, and the analytical solution is not available in general. Then,
we apply a unitary transformation $U_\textrm{JC}=\exp[\frac{i}{2g\sqrt{\eta}}\left(g^{-2}x\tau_y+p\tau_x\right)]$
to Eq.~(\ref{eq:1:03}) so that a transformed Hamiltonian $U_\textrm{JC}^\dagger \bar H_\textrm{JC}(\alpha_g)
U_\textrm{JC}$ is free of  coupling terms between spin subspaces $\mathcal{H}_{\bar\down}$ and $\mathcal{H}_{\bar\up}$. Finally, a projection onto $\mathcal{H}_{\bar\down}$, that is,
$\bra{\bar \down}U^\dagger \bar H_\textrm{JC}(\alpha_g) U\ket{\bar\down}$, leads to the low-energy
effective Hamiltonian of JC model in the superradiant phase,
\begin{align}
    \label{eq:1:04}
    \bar H_{\textrm{JC}}^{\textrm{sp}}=&\frac{\omega_0}{4}(1-g^{-4})x^2+E_G^\textrm{sp}(g),
\end{align}
Here the ground state energy $E_G^\textrm{sp}(g)=-\Omega(g^2+g^{-2})/4$, leading to a discontinuity
in the second derivative of $E_G$ at $g=1$, locating a second order QPT.

\begin{figure}[t]
\centering
\includegraphics[width=0.7\linewidth,angle=0]{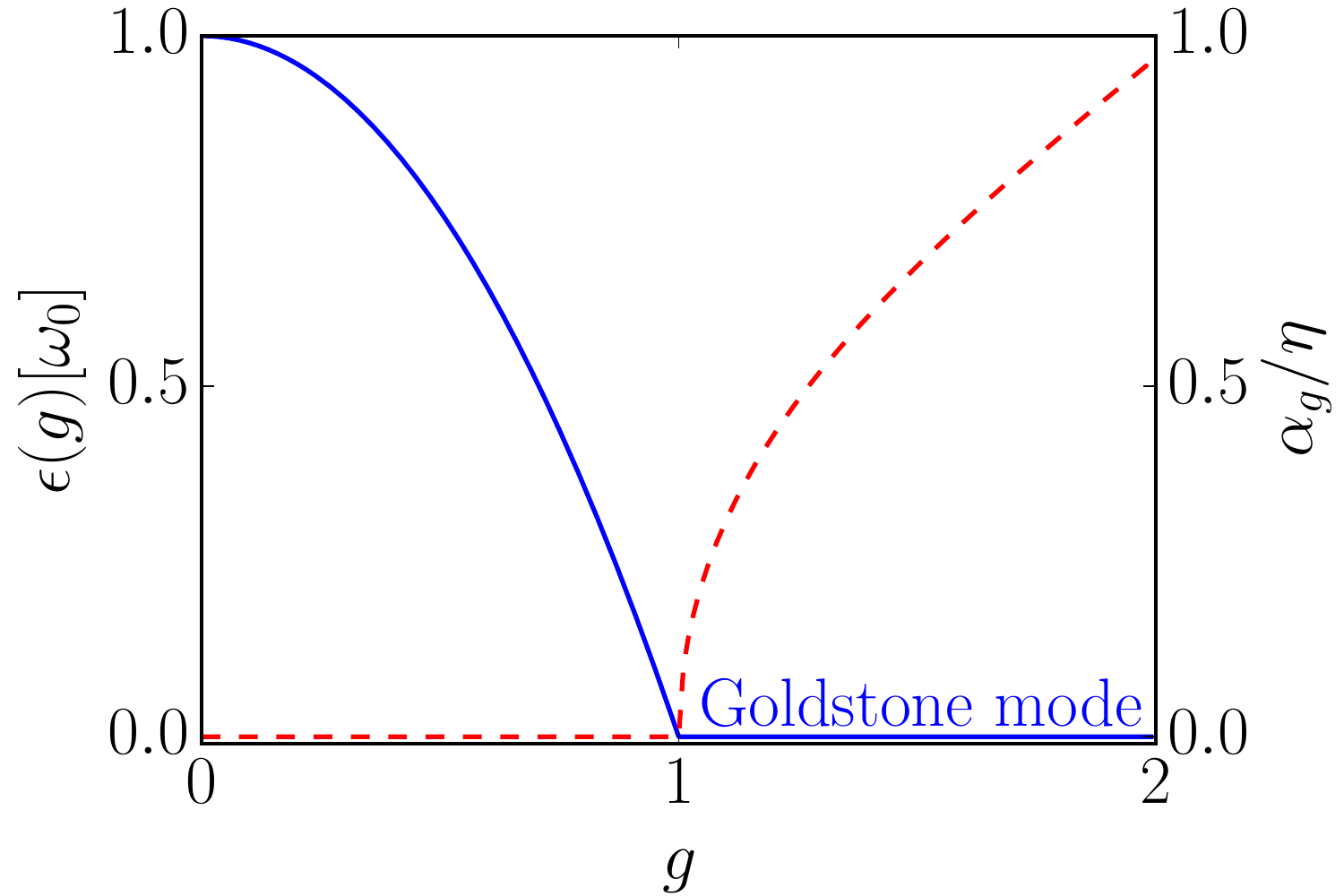}
\caption{QPT of the JC model. The excitation energy $\epsilon(g)$ (left, blue-solid) and the ground state coherence $\alpha_g/\eta$ of the cavity field (right, red-dashed) in the $\eta\rightarrow\infty$ limit. For $g>1$, the $U(1)$ symmetry of the JC model is broken, leading to a Goldstone mode and a non-zero coherence.}
\label{fig:2}
\end{figure}

Interestingly, the effective Hamiltonian is quadratic only in $x$ quadrature of the cavity
field, while $p$ quadrature does not appear in the Hamiltonian in the $\eta\rightarrow\infty$
limit \footnote{Note that for $\bar H_\textrm{JC}(\alpha_g,\theta)$, which includes the phase
factor $\theta$, it is the $x(\theta)=e^{-i\theta a^\dagger a} x e^{i\theta a^\dagger a}$
quadrature that would appear in the effective Hamiltonian~\cite{sup}.}. The ground state
of $\bar H_{\textrm{JC}}^{\textrm{sp}}$ is an eigenstate of $x$ quadrature, which is an
infinitely squeezed vacuum, whose major axis is the $p$ quadrature, i.e., $\ket{r\rightarrow\infty}
= \lim_{r\rightarrow\infty}\mathcal{S}[r]\ket{0}$ with $\mathcal{S}[r]=\exp[-\frac{r}{2}(a^{\dagger2}
-a^2)]$. Going back to the original basis, the ground state of JC model is $\ket{\Psi_G^\textrm{sp}(\theta)}
= e^{i\theta a^\dagger a}\mathcal{D}[\alpha_g]\mathcal{S}[r\rightarrow\infty]\ket{0}$ for
$\theta\in[0,2\pi]$. Since any choices of the phase $\theta$ of the displacement $\alpha$
lead to an identical spectrum, the ground states are also infinitely degenerate. The ground
state of the JC Hamiltonian for the superradiant phase is therefore an infinitely squeezed
photon condensate, whose renormalized photon occupation number is $n_G/\eta=(g^2-g^{-2})/4$. Moreover, the
$U(1)$ symmetry is spontaneously broken, as it is evident from a non-zero spontaneous coherence
$\braket{a}/\eta \equiv\bra{\Psi_G^\textrm{sp}(\theta)} a\ket{\Psi_G^\textrm{sp}(\theta)}/\eta
= e^{i\theta}\sqrt{(g^2-g^{-2})/4}$, which is an order parameter of the QPT in the JC model
[Fig.~\ref{fig:2}].

We note that the critical behaviors described here, diverging ground state energy, squeezing
and spontaneous coherence, arise only in the limit of $\eta\rightarrow\infty$ as the QPT. For
any \textit{finite} $\eta$, the ground state of the JC model has a finite energy with a
definite number of $\braket{N}_\textrm{tot}$ for any value of $g$; moreover, by the symmetry,
the coherence of the ground state $\braket{a}$ and the squeezing of the ground state is always
zero. We remind that this is exactly analogous with the fact that a model that undergoes a
QPT in the $N\rightarrow\infty$ limit restores analytical behaviors for any finite values
of $N$~\cite{Sachdev:2011uj,Fisher:1972tv, Botet:1983bf}.

Because the Eq.~(\ref{eq:1:04}) is quadratic in only one quadrature without the conjugate
variable appearing in the Hamiltonian, the excitation spectrum is gapless [Fig.~\ref{fig:2}];
that is, it takes infinitesimally small energy to excite the system from the ground state.
This gapless excitation is a well-known consequence of spontaneous symmetry breaking of
continuous $U(1)$ symmetry and is often called as a Goldstone mode~\cite{Goldstone:1962ff}.
We note that the effective photon number potential shown in Fig.~\ref{fig:1} (b) or the
mean-field energy of the JC model~\cite{sup} assumes the form of the mexican-hat potential
in a phase space of the cavity field $a$; therefore, the appearance of the Goldstone mode
can be intuitively understood from the fact that  the excitation along the circle of the
potential minima does not cost any energy. Finally, the vanishing spectral gap near the
critical point gives rise to a critical exponent, $\epsilon(g)\propto|g-1|^\alpha$
with $\alpha=1$, which differs from $\alpha=\half$ of the Rabi model~\cite{Hwang:2015eq}.

We have shown so far that the JC model, one of the most fundamental in quantum optics,
exhibits a second-order QPT. Our analysis clearly demonstrates that the atom-cavity coupling
controls the number of photons in the ground state, and that a large $\eta$ leads
to a divergence in the photon number of the ground state. We note that $\eta$ plays
precisely the same role in the Jaynes-Cummings model as the number of atoms in the
Tavis-Cummings model~\cite{Tavis:1968fwa}, a $N$-atom generalization of the JC model,
which undergoes the same kind of QPT~\cite{Baksic:2014ih}. Therefore, the fact that arbitrarily
many photons can be created through interaction with other quantum system, regardless of
its size, is the origin of the QPTs in a photonic (phononic) system with finite
components is possible, in contrast to systems with a hard-core bosons or spins which
require infinitely many components to achieve a QPT.

{\it Mott-insulator to Superfluid transition in a finite JC lattice model.---} We now
consider a photonic lattice model with a finite lattice size. We demonstrate that this model
is capable of exhibiting Mott-superfluid type phase transitions away from the conventional
thermodynamic limit of infinite lattice sites. Specifically, we consider the JC lattice model
\cite{Hartmann:2006kv,Greentree:2006jg,Angelakis:2007ho,Hartmann:2008ei}
which describes a one-dimensional lattice of coupled cavities each containing a two-level atom to
realize the JC model, which reads $H_\textrm{JCL}=\sum_{i=1}^{N}H_\textrm{JC,i}+\sum_{i=1}^{N-1}
J(a_ia^\dagger_{i+1} +h.c.)$, where $i$ indicates $i$-th cavity and $H_\textrm{JC,i} =
\omega_0 a_i^\dagger a_i +\frac{\Omega}{2}\sigma_{iz}-\lambda (a_i \sigma_{i+}+a_i^\dagger
\sigma_{i-})$. The model has a global $U(1)$ symmetry due to the conservation of the total
excitation number, $N_\textrm{tot}=\sum_i(a^\dagger_ia_i+\sigma_{i+}\sigma_{i-})$. In the
$N\rightarrow\infty$ limit, it is in general not amenable to exact solutions, neither
analytically nor numerically; therefore its phase diagram, showing the Mott-insulating-superfluid
transition, is often studied based on the mean-field solution~\cite{Greentree:2006jg,Koch:2009hh}.
For finite $N$, the numerially exact calculation shows a crossover from a Mott insulating phase
to a superfluid-like phase, due to the finite-size effect which generally prevents the system
undergoing a true QPT~\cite{Angelakis:2007ho}.

\begin{figure}[t]
\centering
\includegraphics[width=0.49\linewidth,angle=0]{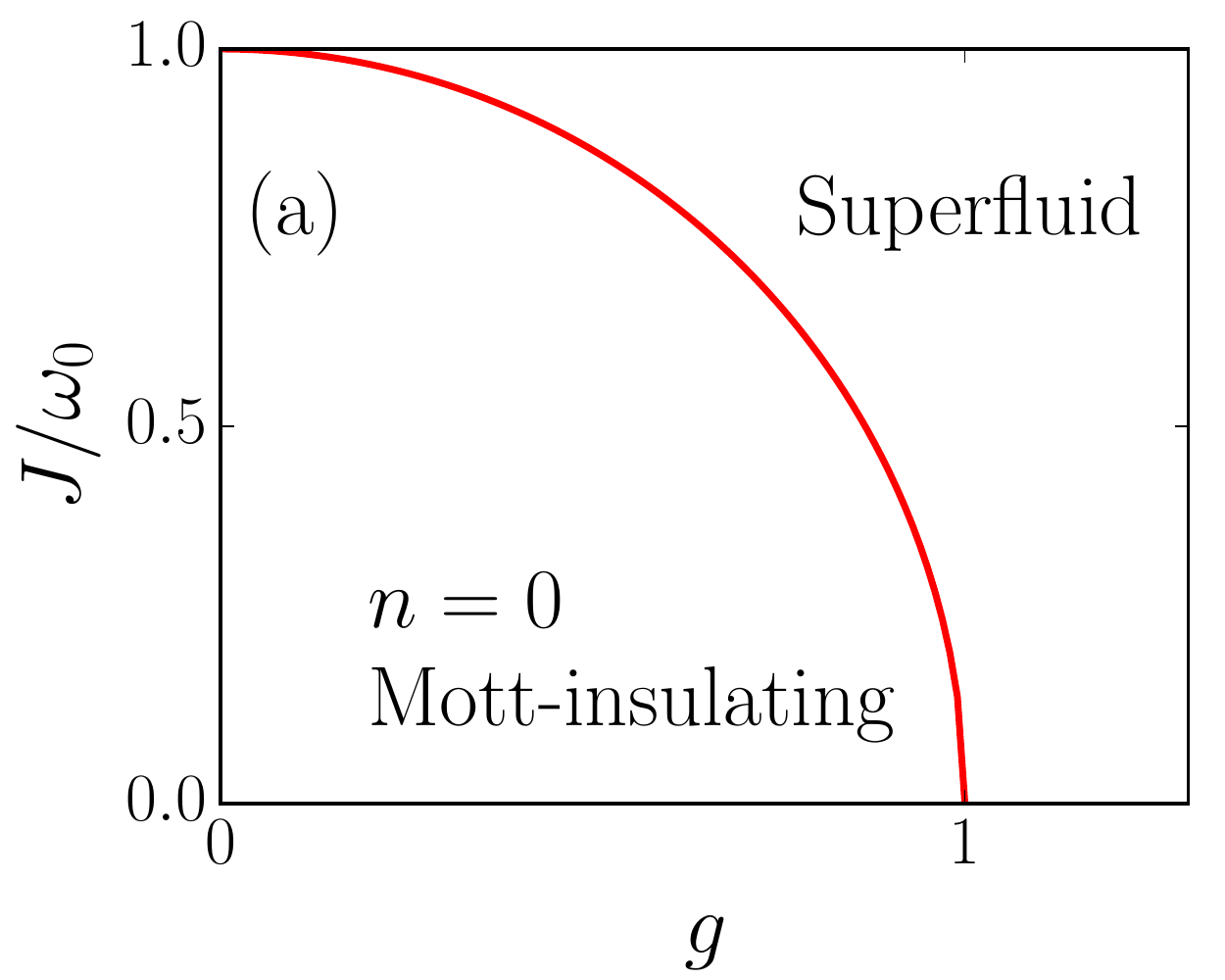}
\includegraphics[width=0.49\linewidth,angle=0]{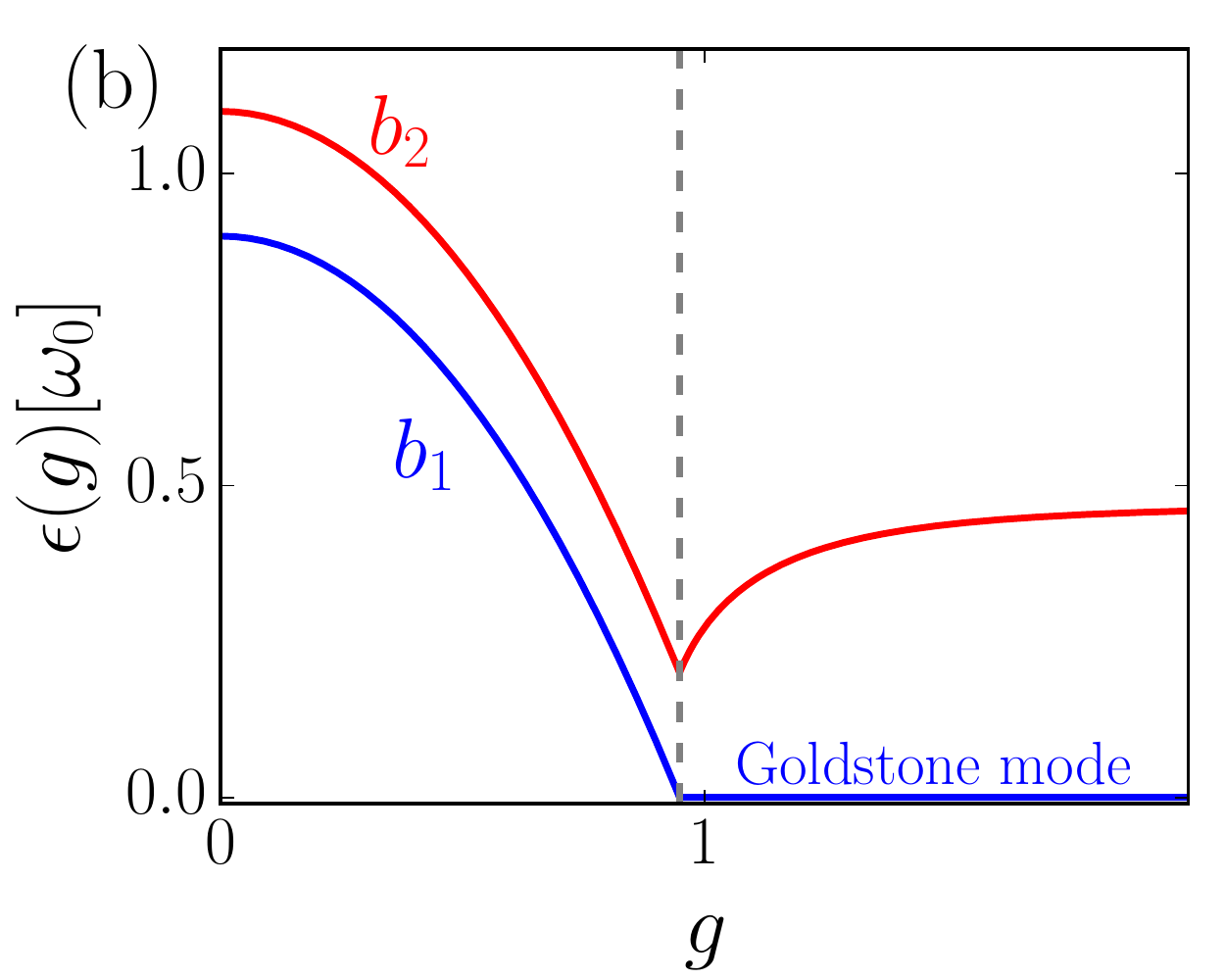}
\caption{JC lattice model. (a) Phase diagram in the $(g,J)$ plane. (b) Excitation energy of the  anti-symmetric ($b_1$) and symmetric ($b_2$) normal mode as a function of $g$ for $J/\omega_0=0.1$. At the critical point, where the $b_1$ mode becomes the Goldstone mode, the first derivative of excitation energy of the $b_2$ mode becomes discontinuous.}
\label{fig:3}.
\end{figure}

We now choose $N=2$, thus called as JC dimer, which is the smallest possible number of sites
for a lattice system, and show that it undergoes a second-order Mott-insulating-superfluid QPT,
in the $\eta\rightarrow\infty$ limit. Note that, unlike some of the previous works \cite{Greentree:2006jg,Koch:2009hh}, we introduce neither a chemical potential term to
fix the number of polaritons nor counter-rotating terms, which has been shown to stabilize
the chemical potential in Ref.~\cite{Schiro:2012hr}; rather, as witnessed in the previous
section, a strong JC-type interaction between the field and the atom itself modulates the number
of polaritons of each cavity. The JC dimer Hamiltonian can be written in in terms of two normal
modes, $b_1=(a_1-a_2)/\sqrt{2}$ and $b_2=(a_1+a_2)/\sqrt{2}$, that is,
\begin{align}
    \label{eq:1:05}
    H_\textrm{JCD}&=(\omega_0-J)b_1^\dagger b_1+(\omega_0+J)b_2^\dagger b_2+\frac{\Omega}{2}\sum_{i=1}^{2}\sigma_{z}\nonumber\\
    &-\frac{\lambda}{\sqrt{2}}(b_1(\sigma_{1+}-\sigma_{2+})+b_2(\sigma_{1+}+\sigma_{2+})+h.c.),
\end{align}
where we assume $J/\omega_0<1$. In the following we treat the two cases $g<g_c$ and $g>g_c$
which lead to different phases separately. To treat the $g<g_c$ case, we first apply a
unitary transformation to $H_\textrm{JCD}$, which decouples the normal modes from the atom,
$U_\textrm{JCD} = \exp[\frac{g}{\sqrt{2\eta}}( b_1(\sigma_{1+}-\sigma_{2+})+b_2(\sigma_{1+}+\sigma_{2+}) - h.c.)]$, and then project the transformed Hamiltonian onto
the subspace of the ground state of two atoms $\ket{\down}_1\ket{\down}_2$~\cite{sup}. We note
that a similar method has been used in Ref.~\cite{Zhu:2013uc} to study the JC lattice in the
dispersive regime. The resulting Hamiltonian is
\begin{align}
    \label{eq:1:06}
    H_\textrm{JCD}^\textrm{Mott}&=\omega_0(1-g^2-\frac{J}{\omega_0})b_1^\dagger b_1+\omega_0(1-g^2+\frac{J}{\omega_0})b_2^\dagger b_2\nonumber\\
    &~-\Omega+\mathcal{O}(\eta^{-\half}),
\end{align}
which becomes exact in the $\eta\rightarrow\infty$ limit. Note that there is a phase boundary $g_c(J)=\sqrt{1-J/\omega_0}$ [Fig.~\ref{fig:3} (a)], on which the spectral gap of the $b_1$
mode vanishes as $\epsilon\propto(g-g_c(J))^\mu$ with $\mu=1$ and beyond which the $b_1$ mode
becomes unstable. As a consequence, Eq.~(\ref{eq:1:06}) is the valid effective Hamiltonian
only for $g<g_c(J)$. In this phase, the ground state is the simple vacuum $|0,\downarrow\rangle$
in the original cavity field basis, $\ket{0,\down}_{1}\ket{0,\down}_{2}$. This corresponds to
a $n=0$ Mott-insulating phase, where each cavity assumes a fixed same number of excitation. The
$b_2$ mode remains to be stable for $g<g_c(J)$.

As we have encountered already in the study of the JC model, the fact that the $b_1$-mode
becomes unstable for $g>g_c(J)$ suggests that it gets occupied by a macroscopic number of photons.
Therefore, it is insightful to look at the mean-field energy of $H_\textrm{JCD}$, which we find
as $E^\textrm{MF}_\textrm{JCD}(\eta,g,J/\omega_0,\beta)/\Omega = g_c^2(J)\eta^{-1}|\beta|^2-\sqrt{1+2g^2\eta^{-1}|\beta|^2}$~\cite{sup}.
It is evident that $E^\textrm{MF}_\textrm{JCD}$ assumes the form of the mexican-hat potential for
$g>g_c(J)$ where the potential minimum occurs at $\beta_1=e^{\theta_1}|\beta_1|$ with $|\beta_1|=\sqrt{\eta/(2g_c^2(J))}\sqrt{(g/g_c(J))^2-(g/g_c(J))^{-2}}$. The mean-field solution predicts a
spontaneously broken-symmetry phase and an appearance of the Goldstone mode. It is also easy
to show that the second derivative of the ground state energy in $g$ become discontinuous at
$g=g_c(J)$, indicating that it is a second order QPT~\cite{sup}.

For $g>g_c(J)$, we obtain a low-energy effective Hamiltonian of the JC dimer using a similar
strategy successfully used for the JC model in the first part of this letter. That is, we first
displace the $b_1$ mode by its mean-field amplitude $\beta_1$, which leads to a new quantization
axis for two atoms and a new atomic state for the ground state~\cite{sup} . Then, just like we
did in the Mott phase, we find a unitary transformation decoupling the normal modes and atoms,
followed by a projection onto the low-energy subspace~\cite{sup}. The resulting effective Hamiltonian
reads
\begin{align}
    \label{eq:1:7}
    \bar H_\textrm{JCD}^\textrm{SF}=&\frac{\omega_0g_c^2}{4}(1-\frac{g_c^4}{g^4}) x_1^2+\frac{J}{2}p_2^2+\frac{\omega_0}{4}\left(1+\frac{J}{\omega_0}-\frac{g_c^6}{g^4}\right)x_2^2
\end{align}
up to the constant ground state energy and $g_c$ here denotes $g_c(J)$.

The two normal modes are decoupled from each other, and the above Hamiltonian is exactly
solvable. First, the $p_1$ quadrature of the $b_1$ mode disappears from the effective
Hamiltonian, as in the case of the JC model in the superradiant phase shown in Eq.~(\ref{eq:1:04}).
Therefore, it immediately follows that the global $U(1)$ symmetry of the JC lattice model
is broken for $g>g_c(J)$. The non-zero coherence of each cavity field $\braket{a_i}\neq0$
marks the onset of the \textit{superfluid} phase and becomes an order parameter. The excitation
spectrum of the $b_1$ mode is gapless, showing that the Goldstone mode correctly emerges as
the low-energy excitation in the broken symmetry phase [Fig.~\ref{fig:3} (b)]. The Hamiltonian
for the $b_2$ mode in Eq.~(\ref{eq:1:7}) can be easily diagonalized to give a harmonic
spectrum with an excitation frequency of $\epsilon_2^{SF}(g)=J\sqrt{2\left(1+\omega_0/J\left(1-g_c^6(J)/g^4\right)\right)}$.
As shown in Fig.~\ref{fig:3} (b), the $b_2$ mode remains gapped for both phases. Interestingly,
the first derivative of $\epsilon_2^\textrm{SF}(g)$ is discontinuous at $g=g_c(J)$. Such a slope discontinuity
of the $b_2$ mode can be potentially used to detect the presence of the Goldstone mode as
suggested in Ref.~\cite{Baksic:2014ih}. The ground state of the $b_2$ mode is a squeezed vacuum,
whose squeezing parameter is given by $\xi=-1/4\ln(\half\left(1+\omega_0/J\left(1-g_c^6(J)/g^4\right)\right)$, which is
zero at $g=g_c$ and gradually increases. The JC-dimer may also serve as testing ground for
the physics of phase interfaces in lattice systems~\cite{Hartmann:2008iu}.

{\it Conclusion--} Unlike massive particles, photons can be created by its interaction
with an atom, as the chemical potential of the photon vanishes~\cite{Landau:1980up}. We
have shown that for an atom with a much larger characteristic frequency than the photon but
strongly coupled to it, it is possible to have a macroscopic photon occupation in the ground
state. This, as we have demonstrated at the hand of the JC models, leads to the emergence of
a QPT in a system composed of finitely many components, photonic modes and atoms. We expect
that our finding here, together with one presented in the Ref.~\cite{Hwang:2015eq}, opens up
an important possibility to study critical phenomena of light and sound, such as QPT, universality,
and dynamics of the QPT, in a fully controlled, small quantum systems including a superconducting
circuits and trapped-ions.

We acknowledge discussions with Inigo Egusquiza. This work was supported by the EU Integrating
Project SIQS, the EU STREPs DIADEMS and EQUAM, the ERC Synergy Grant BioQ and Alexander von Humboldt
Professorship as well as the DFG via the SFB TRR/21 and SPP 1601.

\bibliographystyle{apsrev4-1}
\bibliography{paper}

\end{document}